\documentclass{iopjournal}

\usepackage{xspace}
\usepackage{graphicx}
\usepackage{subcaption}
\usepackage{comment}
\usepackage{soul}
\usepackage[normalem]{ulem}
\usepackage[numbers,sort&compress]{natbib}

\newcommand{\CFf}{\ensuremath{\mathrm{CF}_4}\xspace}

\begin{document}

\title{Non-invasive monitoring of process-relevant plasma parameters by Fiber PROES in capacitively coupled Ar/\CFf plasmas}

\author{Florian Beckfeld$^{1,*}$\orcid{0000-0001-8605-2634}, Maximilian Ryppa$^1$\orcid{0009-0008-8649-6524}, Constantin Neuroth$^1$ \orcid{0009-0006-0775-1407}, Li Wang$^1$\orcid{0000-0002-3106-2779}, Ihor Korolov$^1$\orcid{0000-0003-2384-1243}, and Julian Schulze$^1$\orcid{0000-0001-7929-5734}}

\affil{$^1$Chair of Applied Electrodynamics and Plasma Technology, Department of Electrical Engineering and Information Science, Ruhr University Bochum, D-44780 Bochum, Germany}

\affil{$^*$Author to whom any correspondence should be addressed.}

\email{beckfeld@aept.rub.de}


\begin{abstract}

Monitoring changes of process-relevant plasma parameters, such as the electron density and ion flux to the wafer, is essential for the development and control of plasma processes. However, invasive plasma diagnostics, such as probe measurements, typically cannot be applied to commercial reactors. At the example of a low pressure capacitive radio frequency discharge operated in different mixtures of CF$_4$ and Ar, we demonstrate that changes of such plasma parameters can be monitored non-invasively by phase-resolved optical emission spectroscopy via optical fibers (Fiber PROES), for which the ports are usually available at industrial plasma sources. In this way, the spatio-temporally resolved dynamics of energetic electrons are tracked by observing a selected emission line. By measuring the electron density and ion flux directly via probe and retarding field energy analyzer diagnostics as a function of driving voltage and pressure, changes of these plasma parameters, including hysteresis effects, are found to be correlated with transitions of the electron power absorption mode revealed by Fiber PROES. Such mode transitions cause the electron energy distribution function (EEDF) to change and, thus, affect such plasma parameters. Based on these findings, Fiber PROES can be used as a non-invasive diagnostic for the monitoring and knowledge-based development of plasma processes.

\end{abstract}

\section{Introduction}

Capacitively coupled plasmas (CCPs) operated in electronegative gases, such as \CFf, are widely used in semiconductor manufacturing for plasma etching due to their ability to produce radicals at low temperatures and to bombard surfaces with highly energetic ions \cite{Lieberman2005, Chabert2011, Donnelly2013}. Due to their relevance to applications, the effects of external parameters such as amplitude, frequency, and waveform of the applied voltage, pressure and gas mixture on process-relevant plasma parameters such as the electron density, electron energy distribution function (EEDF), as well as flux-energy distribution functions of ions and radicals at boundary surfaces have been thoroughly investigated \cite{Denpoh1998, Denpoh2000, Georgieva2003, Georgieva2004, Georgieva2005, Georgieva2006, Donko2006, Donko2007, Proshina2010, Schulze2011a, Dittmann2012, Zhang2012, Proshina2015,  Bruneau2015a, Bruneau2016, Wen2022}.

To precisely develop and control a plasma process, access to these parameters through measurements is required. However, many diagnostic methods are invasive, i.e., they need to be in direct contact with the plasma, such as Langmuir probes, or even placed on the substrate, for example retarding field energy analyzers (RFEA). This disturbs the plasma and, thus, such probes typically cannot be used in commercial plasma sources. This creates a demand for non-invasive plasma diagnostics that provide access to process-relevant discharge parameters.

Fundamentally, such plasma parameters are determined by the electron power absorption dynamics and mode, as demonstrated by a variety of previous studies in \CFf plasmas \cite{Denpoh2000, Proshina2010, Schulze2011a, Schulze2011b, Dittmann2012, Zhang2014, Bruneau2015a, Liu2015, Schuengel2016, Brandt2016, Bruneau2016, Liu2018, Brandt2019, Wen2022}. The electron power absorption mode determines the way electrons gain energy from the electric field in the plasma, which substantially influences the EEDF and thereby the dissociation and ionization of the background gas \cite{Wang_2021, Yao_2026}. For example, CCPs have been shown to transition from the $\alpha$-mode, in which electrons are accelerated by ambipolar power absorption at the expanding sheath edges, to the $\gamma$-mode, in which electrons emitted from the electrode surfaces are accelerated inside the sheaths. The latter leads to ionization avalanches that substantially increase the electron density and can be induced by increasing the driving voltage amplitude and/or pressure of the discharge \cite{Belenguer1990, Daksha2019}. Whether the plasma operates in one mode or the other has also been found to depend on the electrode surface material \cite{Daksha2019, Beckfeld2025a}. In electronegative CCPs, such as \CFf discharges, the drift-ambipolar (DA) mode can be dominant. In the DA-mode, electrons gain energy in a bulk drift electric field caused by a low conductivity due to electron attachment, and an ambipolar electric field at the collapsing sheath edges, induced by local maxima of the electron density \cite{Proshina2010, Schulze2011b}. Due to the strong drift field in the bulk, which can enhance the electron power absorption and, therefore, the generation of radicals and ions through dissociation and ionization of the background gas, the DA mode has a strong influence on process-relevant plasma parameters. A transition from the DA- to the $\alpha$-mode  can be induced by, e.g., increasing the applied voltage, or decreasing the pressure in CF$_4$ gas. The presence of the DA-mode has also been found in other electronegative gases, such as $\mathrm{O}_{2}$ \cite{Kuellig2012, Derzsi2016, Gudmundsson2017, Gudmundsson2021}, $\mathrm{SiH}_{4}$ \cite{Kushner1986, Yan2000, Schuengel_2013}, $\mathrm{Cl}_{2}$ \cite{Proto2021}, and $\mathrm{SF}_{6}$ \cite{Gogolides1992}.

The electron power absorption mode is determined by external control parameters, including, but not limited to, driving voltage, frequency, gas pressure, gas mixture, or gap size. Additionally, previous research demonstrated that the same external conditions can result in different electron power absorption modes depending on previous setpoints, i.e., a hysteresis can occur. A hysteresis as a function of the applied voltage has been found in $\mathrm{SiH_{4}}$ and \CFf plasmas \cite{Bohm1991, Wang2024}. In a $\mathrm{CH}_4$ plasma, a hysteresis was observed as a function of the discharge current \cite{Schweigert2004}. In \CFf plasmas, striations caused by ion oscillations have also shown a hysteresis \cite{Liu2019}.

The electron power absorption mode can be measured via phase-resolved optical emission spectroscopy (PROES), where the emission of the plasma at a specific wavelength is recorded with a time resolution of nanoseconds, and then the electron impact excitation rate from the ground state into a specific excited state is calculated from the emission based on a rate equation model \cite{Gans2003, Schulze2010}. Classical PROES requires expensive ICCD cameras and large windows to access the entire plasma volume. These requirements limit its applicability to commercial plasma processes. Recently, a novel approach towards PROES based on connecting a photomultiplier to a plasma reactor via an optical fiber, called Fiber PROES, was proposed, eliminating these limiting requirements \cite{Beckfeld2025}. As this method allows for measuring the nanosecond time-resolved emission of the plasma on a timescale of milliseconds and is minimally invasive through the use of an optical fiber, it is promising for monitoring the electron power absorption mode in applications.

In this work, we demonstrate how Fiber PROES can be correlated with process-relevant plasma parameters by experimentally investigating electron power absorption mode transitions, including hysteresis effects, in a geometrically symmetric CCP operated in \CFf. We correlate Fiber PROES results with direct measurements of the electron density and ion flux across a wide parameter range. These measurements reveal a strong correlation between transitions of the plasma operation mode from the $\alpha$/DA hybrid- to the pure $\alpha$-mode, i.e., such changes of process-relevant plasma parameters can be monitored by Fiber PROES. Based on a single numerical value, i.e., the ratio of two characteristic peaks of the Fiber PROES signal, the mode transition can be identified. Based on these findings, Fiber PROES is demonstrated to be a fast, non-invasive, and easy-to-interpret method by which changes of process-relevant plasma parameters can be monitored during a plasma process. 

\section{Experimental Setup}
\label{sec:methods}
\begin{figure}[h]
 \centering
        \includegraphics[width=\textwidth]{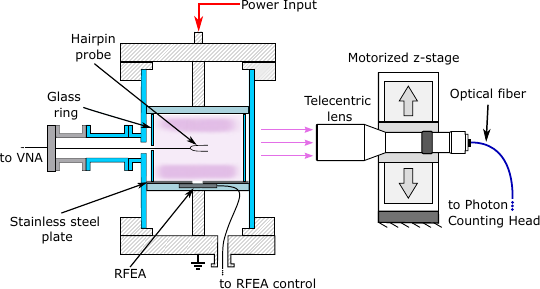}
 \caption{Sketch of the experimental setup.}
\label{setup}
\end{figure}

The experiments in this study are carried out in a geometrically symmetric CCP reactor, shown in figure \ref{setup}. The reactor has been described in detail before in references \cite{Wang2024, Beckfeld2025}. Details on symmetric CCPs can also be found in \cite{Schulenberg2021, Godyak1990}. Hence, only a compact description is provided here.

The reactor consists of two plane-parallel electrodes placed inside a glass cylinder. The diameter of the electrodes is 120\,mm and the electrode gap is 40\,mm. Both electrode surfaces are made of stainless steel. In addition, a glass ring with a thickness of 3\,mm, a diameter of 115\,mm, and a height of the electrode gap is placed inside the chamber to protect the reactor wall from the reactive plasma. The top electrode is powered by a sinusoidal RF voltage at a frequency of 13.56\,MHz while the bottom electrode is grounded. The gases used in this work are Ar at a purity of 6.0 and \CFf at a purity of 5.0, respectively.

Fiber PROES is used to identify the electron power absorption mode. The method is described in detail in \cite{Beckfeld2025}. A telecentric lens (TOB42/11.0-185-V-WN, Vision \& Control GmbH) is mounted on a motorized stage (8MT175-50, Standa Ltd.) and connected to a photon counting head (H16721, Hamamatsu) combined with an optical fiber (M93L02, Thorlabs) to measure the emission intensity of the plasma from two emission lines that result from specific electron transitions, i.e., F I (3p $^{2}P_{3/2}^{0}$ $\rightarrow$ 3s $^{2}P_{3/2}$) at 703.7\,nm, with a lifetime of 17.3\,ns of the upper state, and de-excitation from the Ar 2p$_{1}$ level at 750.4\,nm with a lifetime of 22.2\,ns \cite{NIST_ASD}. To isolate specific emission lines, two bandpass filters with central wavelengths at 703.7\,nm and 750.4\,nm and with a full width at half maximum of 1\,nm are used. Due to their varying photocathode efficiencies at specific wavelengths, two different photon counting heads (H16721-40 and H16721-50) are used. To obtain spatially resolved measurements, the stage was moved along the discharge axis in 0.5\,mm steps. The radio-frequency (RF) period is divided into 37 equal parts for photon counting, which results in a temporal resolution of 2\,ns. After acquiring the nanosecond time-resolved emission, a rate equation model is applied to calculate the spatio-temporally resolved electron-impact excitation rate from the ground state into the upper excited state of the respective transition \cite{Schulze2010, Gans2003}.

\begin{figure}
    \centering
    \includegraphics[width=0.5\textwidth]{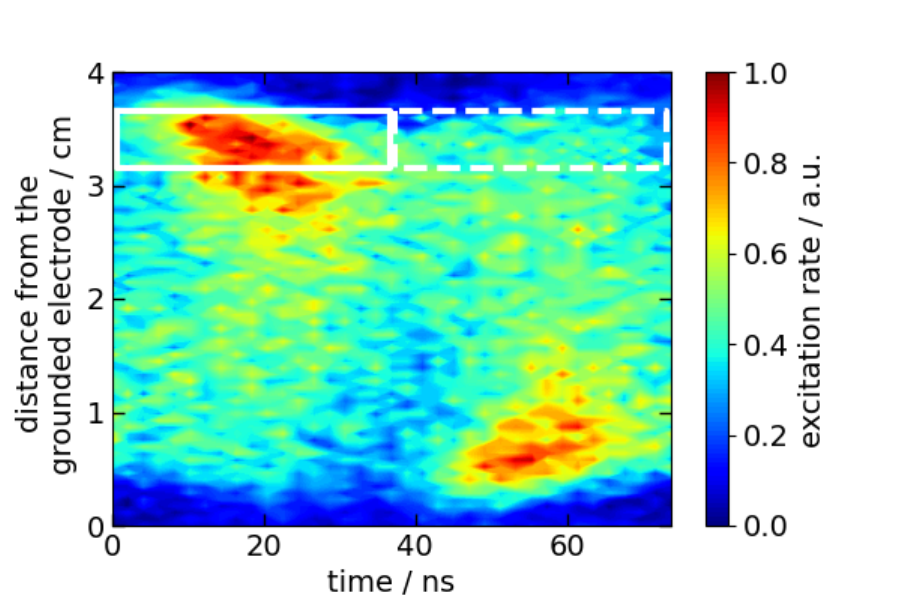}
    \caption{Spatio-temporally resolved electron-impact excitation rate from the ground state to the F I (3p $^{2}P_{3/2}^{0}$) level in a \CFf plasma at 15\,Pa and a driving voltage amplitude of 175\,V. The solid and dashed white rectangles indicate the areas that were used to calculate the intensity ratio of the characteristic DA- and $\alpha$-mode peaks.}
    \label{fig:line_ration_explanation}
\end{figure}

To find a quantitative indicator to determine whether the plasma is in the DA-mode or the $\alpha$-mode, the following algorithm is used on the Fiber PROES measurements: figure \ref{fig:line_ration_explanation} shows an exemplary measurement of the electron-impact excitation rate from the ground state to the F I (3p $^{2}P_{3/2}^{0}$) level in a  \CFf plasma at 15\,Pa and a driving voltage amplitude of 175\,V. By using a peak-finding algorithm, the maximum in the first half of the RF period in the upper half of the discharge reactor is identified. Then, the average intensities within the first and second half-period in a region $\pm$2.5\,mm around this maximum are determined as indicated by the white rectangles in figure \ref{fig:line_ration_explanation}. Subsequently, the ratio $R = I_2/I_1$ is calculated, where $I_1$ is the intensity within the first half period (indicated by the solid rectangle in figure \ref{fig:line_ration_explanation}), and $I_2$ is the intensity of the second half period (indicated by the dashed rectangle in figure \ref{fig:line_ration_explanation}). As the excitation rate within each of these time windows is caused by different electron power absorption mechanisms (here: $\alpha$-mode for I$_1$ and DA-mode for I$_2$), the total electron power absorption mode the plasma is operated in can be determined. A high value of $R$ indicates that the plasma is predominantly operated in the DA-mode due to high excitation during the sheath collapse, while a low value of $R$ indicates a dominant $\alpha$-peak and, thus, that the plasma is in the $\alpha$-mode. Generally, this value is expected to change with the emission line that is monitored, since electrons at different energies are probed.

To measure the ion flux, an RFEA (Semion, Impedans) is placed on the grounded electrode \cite{Gahan2008}. The RFEA holder is made of aluminum with a cutout with a diameter of 80\,mm and a depth of 5\,mm to place the RFEA inside the chamber with minimal disturbance. To maintain an even surface and similar surface conditions at both electrodes, a stainless steel plate with a hole of a diameter of 10 \,mm in the middle is placed on top of this holder. The ion flux measured with the sensor is not calibrated, hence it is given in arbitrary units \cite{Ries2021, Baloniak2010}.

A hairpin probe is used to measure the electron density in the discharge center, as shown in figure \ref{setup} \cite{Stenzel1976}. The resonator is designed as a reflection type probe with a floating probe tip \cite{Piejak2004, Sands2007}. The probe tip is made of silver, the width $w$ of the resonator is 15\,mm, the length $l$ is 35\,mm, and the diameter of the wire is 0.1\,mm. A 50\,$\Omega$ coaxial cable is connected to a vector network analyzer (PicoVNA 108, Pico Technology) to measure the reflection of the probe. Floating steel meshes are placed in the side flanges of the chamber to keep the discharge confined between the electrodes \cite{Schulenberg2021}. The coaxial cable is guided through a ceramic tube (not shown in figure \ref{setup}) to prevent contact with the meshes. From the measured resonance frequency in the presence of a plasma, $f_\mathrm{r}$, and the vacuum resonance frequency, $f_{0}$, the electron density, $n_\mathrm{e}$, can be determined via 

\begin{equation}
    n_\mathrm{e} = \frac{f_\mathrm{r}^2 - f_0^2}{0.81}, 
    \label{eq:nehairpin}
\end{equation}
where $n_\mathrm{e}$ is given in units of 10$^{16}$ $\mathrm{m^{-3}}$, and $f_\mathrm{r}$ and $f_{0}$ are given in GHz \cite{Piejak2005}.

The electron density measurements in this work are presented without a correction for the presence of a sheath around the probe tip. This is done for several reasons: Passive sheath compensation is not feasible because it would require additional components inside the plasma reactor, significantly perturbing the discharge \cite{Godyak1992}. Active sheath compensation analogous to approaches developed for Langmuir probes \cite{Braithwaite1987, Dyson2000} has not yet been established for hairpin probes and is beyond the scope of this work. Previous works have applied a correction factor based on the assumption of a cylindrical sheath around the probe \cite{Piejak2004}, but as the plasma bulk changes from electronegative to electropositive at the mode transition from the DA- to the $\alpha$-mode, a different sheath model would be required for either mode, which is not available at the moment. Nevertheless, the results from other diagnostics presented in this work, together with a comparison of the uncorrected values of $n_\mathrm{e}$ with previous numerical simulations \cite{Wang2024}, provide support for the conclusion that the measured trends remain physically meaningful despite the absence of a sheath correction.

\section{Results}

\vspace{0.3cm}

\subsection{Pressure Variation}

\vspace{0.3cm}

\begin{figure}[htbp]
 \centering
    \includegraphics[width=\textwidth]{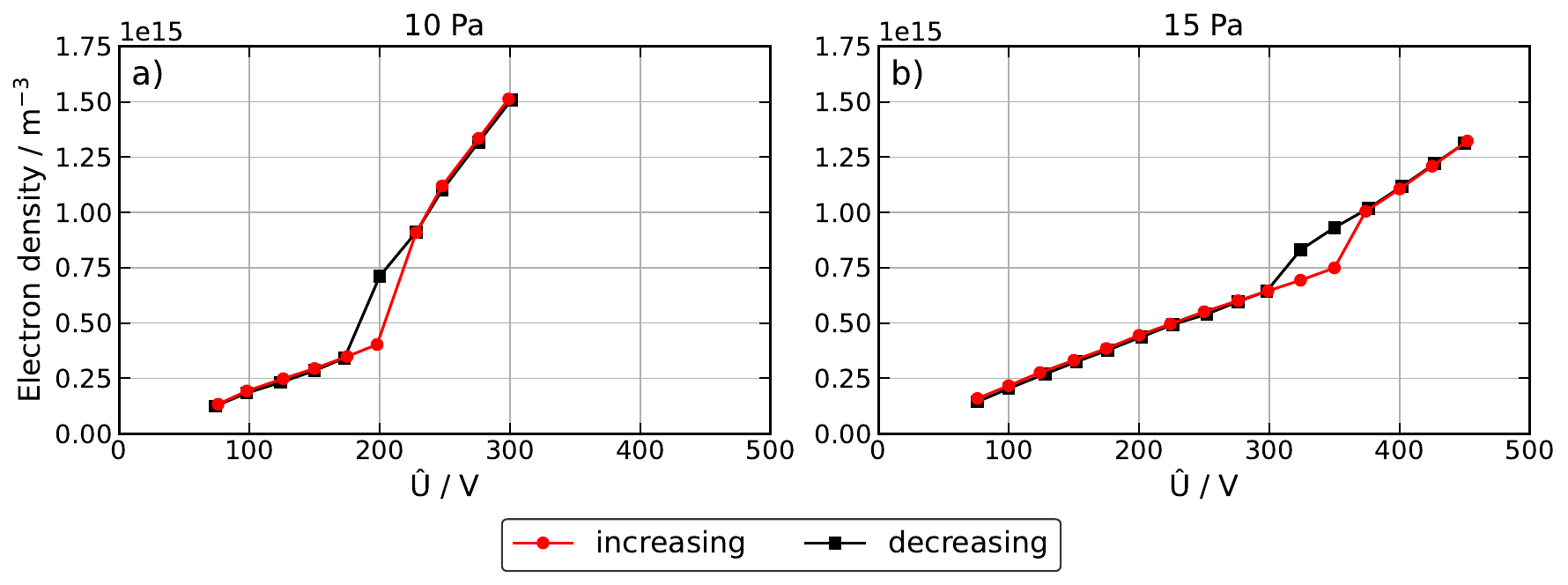}
 \caption{Electron density at the discharge center in \CFf plasmas as a function of the driving voltage amplitude at pressures of 10\,Pa a) and 15\,Pa b) for an increasing and decreasing voltage sweep, respectively.}
\label{fig:electron_density_15&10Pa}
\end{figure}

Figure \ref{fig:electron_density_15&10Pa} a) shows the electron density measured by a hairpin probe in the center of the discharge at a pressure of 10\,Pa as a function of the driving voltage amplitude applied to the discharge. The discharge was initially generated at a voltage amplitude of 75\,V, then the voltage was increased in 25\,V steps up to 450\,V. From there, the voltage was then decreased back to an amplitude of 75\,V in 25\,V steps.

When the voltage is increased from a low value,  the electron density increases linearly with the applied voltage up to a voltage amplitude of 200\,V. At this threshold voltage, the electron density shows a jump-like increase by more than 100\% induced by increasing the voltage by 25 V. Beyond this transition point, it increases at a steeper slope. When the voltage is decreased, the electron density shows a hysteresis, i.e., it returns to the density values of the increasing branch at a lower threshold voltage.

The same behavior is found at a pressure of 15\,Pa (figure \ref{fig:electron_density_15&10Pa} b)). First, when increasing the voltage, the electron density increases slowly. Then, at a threshold voltage of 350\,V, it increases abruptly and continues to increase with a steeper slope. When the voltage is decreased, the electron density shows a hysteresis.

\begin{figure}[htbp]
    \centering
    \includegraphics[width=\linewidth]{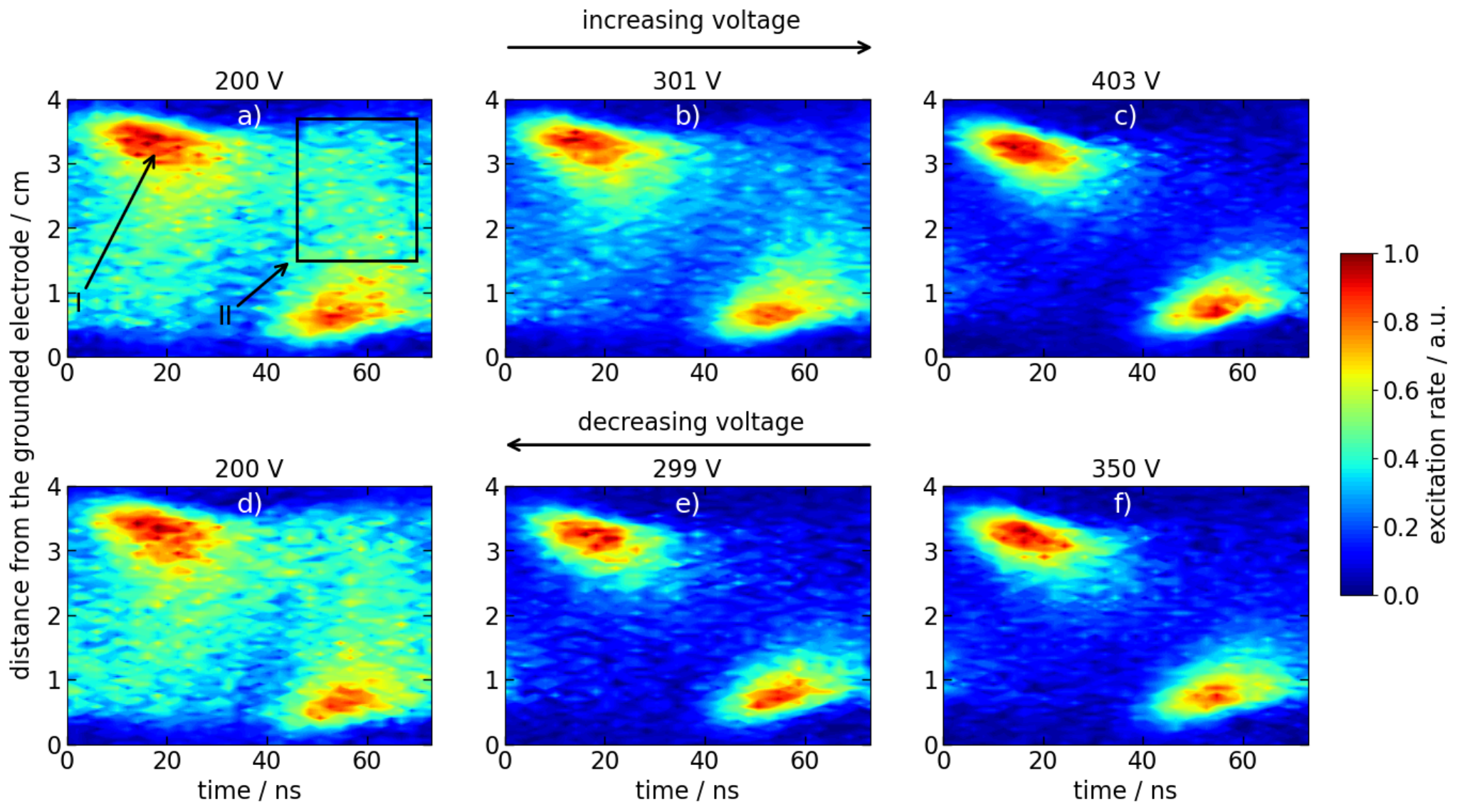}
    \caption{Spatio-temporally resolved electron-impact excitation rate from the ground state into the F I (3p $^{2}P_{3/2}^{0}$) level in a \CFf plasma at 15 Pa. The top row shows measurements for an increasing voltage, while the bottom row shows results for a decreasing voltage. Each plot is normalized to its maximum. Zone I indicates the $\alpha$-peak, while Zone II indicates excitation by the bulk drift electric field.}
    \label{fig:FP_15Pa_CF4}
\end{figure}

The reason for the observed behavior of the electron density is a transition of the electron power absorption mode from a hybrid $\alpha$/DA-mode to the $\alpha$-mode, as can be seen in the Fiber PROES measurements. Figure \ref{fig:FP_15Pa_CF4} shows Fiber PROES measurements of the electron-impact excitation rate from the ground state into the F I (3p $^{2}P_{3/2}^{0}$) level in a \CFf plasma at 15\,Pa for an increasing voltage (top row) and a decreasing voltage (bottom row). The x-axis shows one RF period of 74\,ns, while the y-axis shows the electrode gap of 4\,cm.

In figure \ref{fig:FP_15Pa_CF4} a), which shows a measurement at 200\,V driving voltage amplitude, the plasma is in a hybrid $\alpha$/DA electron power absorption mode \cite{Belenguer1990, Proshina2010, Schulze2011b}. This is indicated by excitation peaks near the electrodes caused by the expansion of the sheaths ('$\alpha$-peak', Zone I in figure \ref{fig:FP_15Pa_CF4} a)), as well as significant excitation in the plasma bulk (Zone II in figure \ref{fig:FP_15Pa_CF4} a)), caused by electron power absorption by the drift-field in the bulk due to the low conductivity caused by the electronegativity of the plasma. When the voltage is increased (figure \ref{fig:FP_15Pa_CF4} b)), the $\alpha$-peaks get stronger compared to the excitation due to the drift field electron power absorption. Increasing the voltage to 403\,V (figure \ref{fig:FP_15Pa_CF4} c)), the excitation within the plasma bulk caused by the drift field vanishes, indicating a transition from the hybrid $\alpha$/DA-mode to the $\alpha$-mode. When the voltage is decreased from 403\,V (figure \ref{fig:FP_15Pa_CF4} d) - f)), the plasma remains in the $\alpha$-mode, even when the voltage is reduced to around 300 V, where the plasma was in the hybrid $\alpha$/DA-mode for a voltage that was increased from a lower value. Clearly, the excitation rate in Zone II is close to zero in figure \ref{fig:FP_15Pa_CF4} e), indicating that the drift-field induced excitation is weak, while this is not the case in figure \ref{fig:FP_15Pa_CF4} b). By further decreasing the driving voltage, the electron power absorption mode transitions back to the hybrid $\alpha$/DA-mode (figure \ref{fig:FP_15Pa_CF4} d).
  
\begin{figure}[htbp]
\centering
    \includegraphics[width=\textwidth]{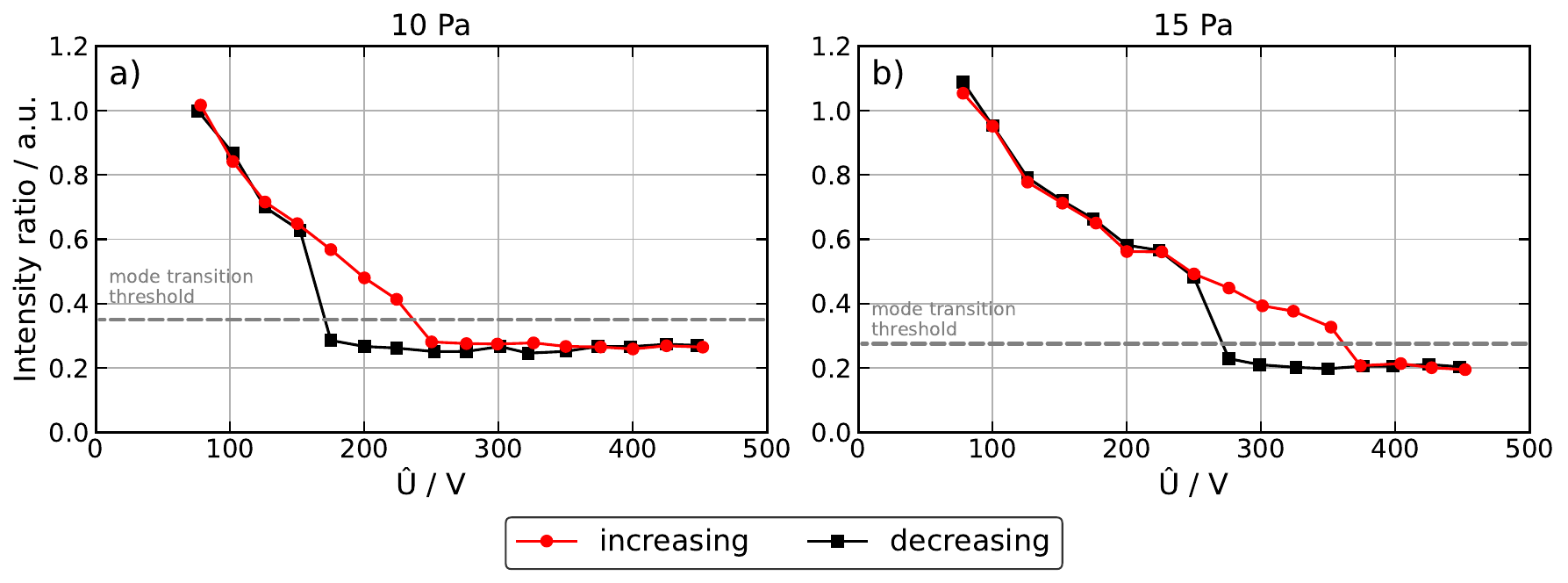}
    \caption{Ratio of the intensities averaged over the regions of interest indicated in figure \ref{fig:line_ration_explanation}  within the first and second half of the RF period as a function of the driving voltage amplitude obtained based on the measured electron-impact excitation rate from the ground state to the F I (3p $^{2}P_{3/2}^{0}$) level and the method described in section \ref{sec:methods}. The plots show results for pressures of 10\,Pa (a) and 15\,Pa (b), respectively.}
    \label{fig:Intensity_ratio_10&15Pa}
\end{figure}

The transition in the electron power absorption mode can be observed more clearly when looking at the intensity ratio determined from the Fiber PROES measurements as described in section \ref{sec:methods}, which is shown in figure \ref{fig:Intensity_ratio_10&15Pa}. When increasing the voltage from a low value, the intensity ratio is $\approx$ 1, due to the electron-impact excitation caused by the drift field being equally strong as the electron-impact excitation due to the expanding sheath. As the $\alpha$-peak gets stronger, the intensity ratio decreases to a value of $\approx$ 0.2 at a voltage of 350\,V (figure \ref{fig:Intensity_ratio_10&15Pa} b)), where the plasma transitions to the $\alpha$-mode, as observed in figure \ref{fig:FP_15Pa_CF4} c). The intensity ratio then remains constant as there is only background excitation in the second half of the RF period. When decreasing the voltage from a high value, the intensity shows a jump-like increase at a certain threshold voltage, which shows that the electron power absorption mode of the plasma returns back to the hybrid $\alpha$/DA-mode. As indicated by the dashed line in figure \ref{fig:Intensity_ratio_10&15Pa}, a threshold value in the intensity ratio can be defined to determine the electron power absorption mode, based on the value where the intensity ratio does not change as a function of voltage. This value is not the same for each pressure, since the emission from the plasma gets weaker compared to the background at 10 \,Pa. Therefore, a general threshold value cannot be determined. In applications, the threshold at which the mode transition takes place can be determined by performing a calibration where a voltage sweep from a low to a high voltage is done and the threshold is set between the value where the intensity ratio stays constant and the value before that, e.g., between 225\,V and 250\,V in figure \ref{fig:Intensity_ratio_10&15Pa} a).

The behavior of the electron density shown in figure \ref{fig:electron_density_15&10Pa} can be connected to the electron power absorption mode transition (figure \ref{fig:FP_15Pa_CF4} and \ref{fig:Intensity_ratio_10&15Pa}) in the following way: in plasmas containing fluorine, electron attachment causes the creation of F$^{-}$ ions, depleting the electron density in the bulk \cite{Denpoh2000}. A drift field in the plasma bulk forms due to the low conductivity \cite{Proshina2010, Schulze2011b}. The electrons are accelerated within the drift electric field, which benefits dissociative electron attachment to \CFf molecules, the cross section of which is high between 5 - 10\,eV \cite{Kurihara2000}. This results in a high F$^{-}$ density and a low electron density in the plasma bulk, as seen in figure \ref{fig:electron_density_15&10Pa} \cite{Liu2015, Wang2024}. A hybrid $\alpha$/DA-mode is formed. When the voltage is moderately increased, the bulk field and mean electron energy remain high so that the production of F$^{-}$ continues to consume electrons, which therefore prevents a significant increase of the electron density. When the voltage is further increased, at a certain threshold, the bulk electric field becomes too weak to drive dissociative electron attachment to F atoms, causing a surge in electron density observed in figure \ref{fig:electron_density_15&10Pa} \cite{Liu2015}. This, in turn, causes the strength of the bulk electric field to significantly decrease, which is then visible as a transition to the $\alpha$-mode (figure \ref{fig:FP_15Pa_CF4} c)). When the voltage is subsequently decreased with the plasma in the $\alpha$-mode, the initial condition of the discharge is characterized by a weak bulk electric field, which is only enhanced when the voltage is decreased to a lower value, due to a decreasing bulk conductivity. When the electric field reaches a critical strength, it will drive the formation of F$^{-}$ ions through electron attachment, causing the electron power absorption mode to transition to the hybrid $\alpha$/DA-mode \cite{Wang2024}. The difference in the threshold voltage for the mode transition between figure \ref{fig:electron_density_15&10Pa} and figure \ref{fig:Intensity_ratio_10&15Pa} can be explained by the presence of the hairpin probe, which slightly perturbs the discharge. The shift of the mode transition to lower voltages with decreasing pressure is explained by the fact that the attachment rate decreases with pressure; simultaneously, the electron mean free path increases, hence they are accelerated to higher energies, increasing their potential to ionize \cite{Denpoh2000}. By this, the threshold voltage for the mode transition is lower.

\begin{figure}[htbp]
\centering
    \includegraphics[width=\textwidth]{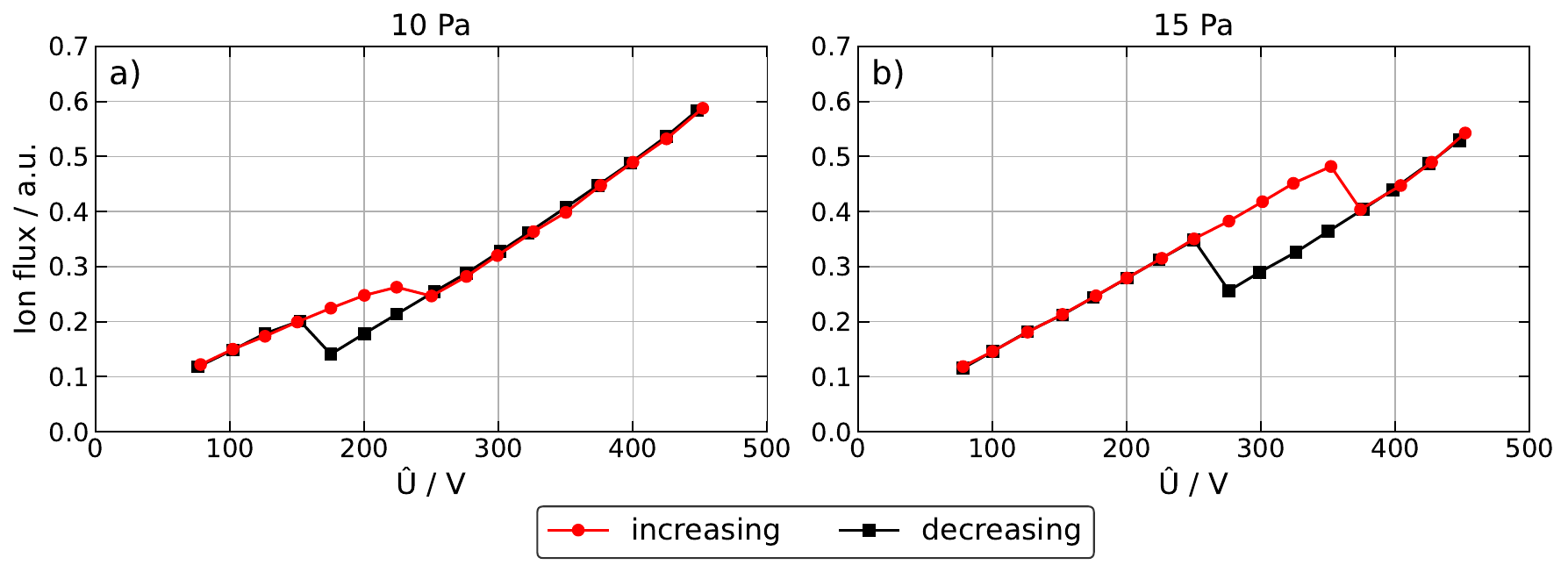}
    \caption{Measurements of the ion flux at the grounded electrode in a \CFf plasma at 10\,Pa (a) and 15\,Pa (b), for an increasing and decreasing voltage sweep, respectively.}
    \label{fig:Ion_Flux_10&15Pa}
\end{figure}

Figure \ref{fig:Ion_Flux_10&15Pa} shows the ion flux at the grounded electrode for increasing and decreasing voltage sweeps in a  \CFf plasma at 10\,Pa and 15\,Pa, respectively. Figure \ref{fig:Ion_Flux_10&15Pa} a), shows that the ion flux increases linearly starting from a low voltage, then shows a step-wise decrease at a voltage of 250\,V, and then continues to increase at a linear rate. As the voltage is decreased from a high value, the ion flux is lower in the decreasing branch than in the increasing branch within a window between 150\,V and 250\,V, by $\approx$ 35\%. The same behavior is observed at a pressure of 15\,Pa, with the threshold values for the stepwise change of the ion flux moving to higher voltages, and the region in which the ion flux in the decreasing branch is lower increasing. The changes in the trend of the ion flux, as well as the regions in which the ion flux is different, correspond to changes in the electron power absorption observed in figure \ref{fig:Intensity_ratio_10&15Pa}.

The behavior of the ion flux is explained as follows: when the plasma is in the hybrid $\alpha$/DA-mode, the electric field in the bulk causes strong dissociation and ionization of the background gas, leading to a high ion flux. As the electric field in the plasma bulk vanishes at the transition to the $\alpha$-mode, fewer molecules are ionized, causing a decrease in the ion flux. When the plasma transitions back to the hybrid $\alpha$/DA-mode when decreasing the voltage, the recurring drift field in the bulk causes the ion flux to increase again. The trends in the measurements of the electron density and the ion flux observed in the experiments confirm previous computational findings \cite{Wang2024}.

Overall, these results show that changes in process-relevant plasma parameters are correlated with the electron power absorption mode and its transition induced by changes of external control parameters. Monitoring the spatio-temporally resolved electron impact excitation dynamics via Fiber PROES is, thus, sufficient to track changes thereof. This can be done non-invasively via a fiber access point, typically available at commercial plasma reactors. Spatial resolution of the PROES measurements is not necessarily required, but monitoring the plasma at a distance from the electrode roughly corresponding to the maximum local sheath width suffices.

\subsection{Argon Admixture}

\vspace{0.3cm}

\begin{figure}[h]
 \centering
    \includegraphics[width=\textwidth]{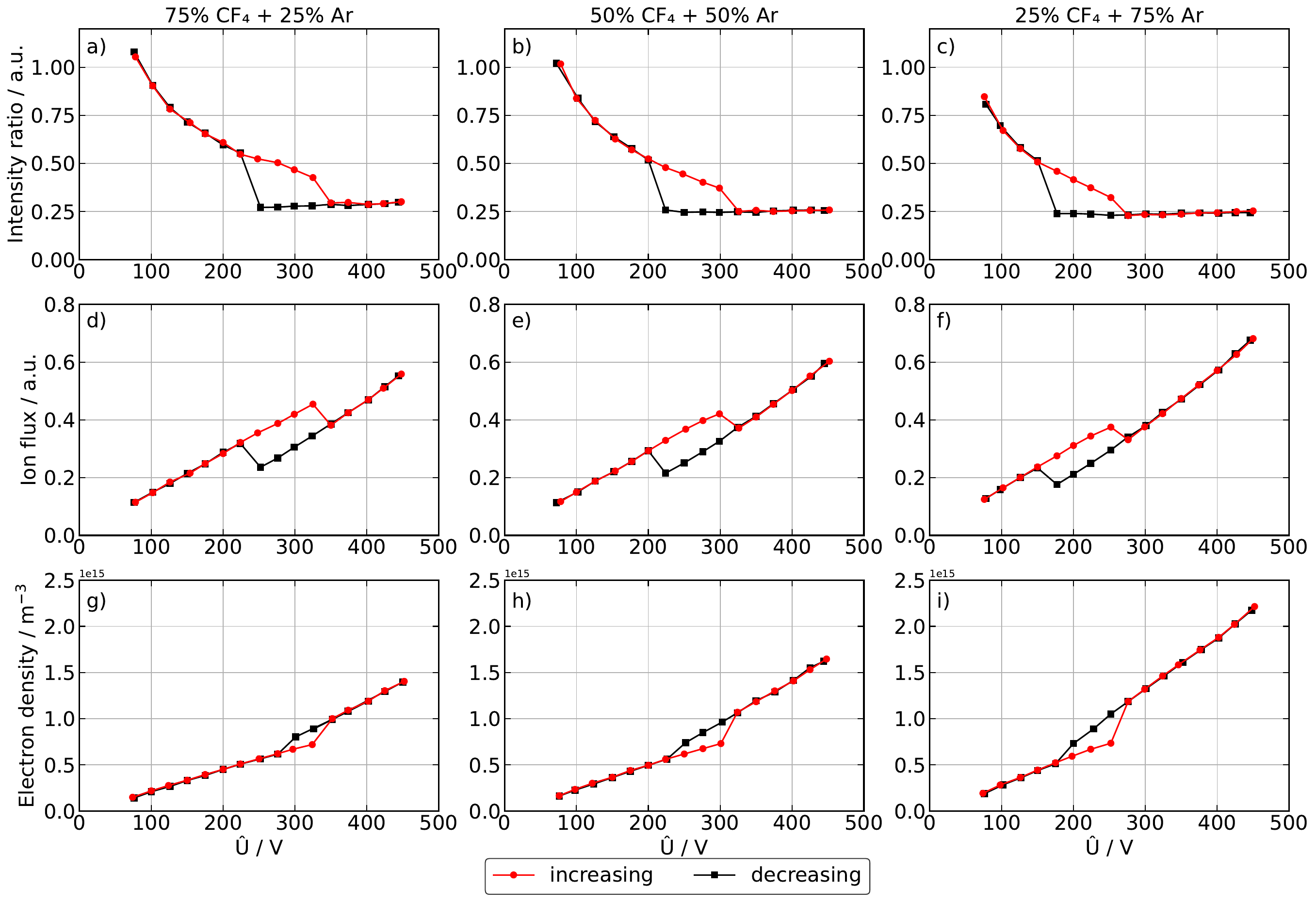}
 \caption{Ratio of the intensities averaged over the regions of interest indicated in figure \ref{fig:line_ration_explanation}  within the first and second half of the RF period of the measured electron-impact excitation rate from the ground state to the Ar 2p$_{1}$ level (first row), ion flux at the grounded electrode (second row), and electron density in the center of the discharge (third row) as a function of the driving voltage amplitude. The pressure is 15\,Pa for all plots, while the argon admixture was set to 25\% (left column), 50\% (middle column), and 75\% (right column). Each plot shows a measurement obtained for an increasing and a decreasing voltage sweep.}
\label{fig:megaplot}
\end{figure}

To expand the study to gas mixtures that are more relevant to plasma processing, the study was continued with \CFf plasmas including various admixtures of Ar. In figure \ref{fig:megaplot}, measurement results in the presence of an argon admixture of 25\% (left column), 50\% (middle column), and 75\% (right column) are shown. The first row (figure \ref{fig:megaplot} a) - c)) shows the intensity ratio determined from Fiber PROES measurements of the electron-impact excitation from the ground state to the Ar 2p$_{1}$ level as explained in section \ref{sec:methods}, the second row (figure \ref{fig:megaplot} d) - f)) shows measurements of the ion flux to the grounded electrode, and the third row (figure \ref{fig:megaplot} g) - i)) shows the electron density measured at the center of the discharge as a function of the driving voltage amplitude.\\
Figure \ref{fig:megaplot} a) shows that the plasma transitions from the hybrid $\alpha$/DA-mode to the $\alpha$-mode at a voltage of 325\,V for an increasing driving voltage and an Ar admixture of 25 \%, which is indicated by the intensity ratio reaching a plateau of $\approx$ 0.25. For an Ar admixture of 50 \% (figure \ref{fig:Intensity_ratio_10&15Pa} b)), the threshold for the mode transition is 25\,V lower. Similar to pure \CFf discharges, there is a hysteresis in the electron power absorption mode transition, with the plasma transitioning back to the hybrid $\alpha$/DA-mode at 250\,V amplitude for 25 \% Ar admixture. When increasing the argon content to 50\% (figure \ref{fig:megaplot} b)), the electron power absorption mode transition in the increasing branch occurs at 325\,V and at 200\,V in the decreasing branch, while at an argon content of 75\% (figure \ref{fig:megaplot} c)) the mode transitions occur at 275\,V (increasing) and 150\,V (decreasing), respectively.\\ 
The changes in the electron power absorption mode influence the ion flux and the electron density in a similar way as it was observed in pure \CFf, i.e., the transition from the hybrid $\alpha$/DA-mode to the $\alpha$-mode leads to a step-wise decrease in the ion flux due to the lack of ionization caused by electron acceleration by the drift field in the bulk. Simultaneously, the electron density shows a step-wise increase at the mode transition as fewer electrons are attached to F atoms. Overall, it can be seen that ion flux and electron density increase with increasing argon content.\\ 
The observed shift in the threshold voltage can be explained by the increased electron density due to the admixture of argon. Admixing argon to a \CFf plasma makes the discharge less electronegative. Thus, the bulk electric field is attenuated. This can also be observed when comparing the intensity ratios at a low applied voltage for different argon admixtures (figure \ref{fig:megaplot} a) - c)). At the lowest applied voltage of 75\,V, the intensity ratio is slightly above 1 for an argon admixture of 25\%, while it is only $\approx$ 0.8 for 75\% admixture. Therefore, as the increased electron density reduces the strength of the bulk electric field, the threshold at which electron attachment becomes ineffective is reached at a lower applied voltage, leading to a mode transition at a lower threshold voltage. The shift in mode transition by admixing argon confirms the explanation of the observed hysteresis effect and matches computational findings in which the $\gamma$-coefficient was varied \cite{Wang2024}.\\

\section{Conclusions}

\vspace{0.3cm}

Electron power absorption mode transitions in capacitively coupled plasmas operated in \CFf and Ar/\CFf mixtures were investigated experimentally by Fiber PROES as a function of the driving voltage amplitude and pressure. These results were correlated with measurements of the electron density and the ion flux to the grounded electrode. Based on this correlation, Fiber PROES was demonstrated to be able to detect transitions in the electron power absorption mode, which cause significant changes in process-relevant plasma parameters. Thus, Fiber PROES can be used to monitor such process-relevant plasma parameters non-invasively. 

By using the intensity ratio method presented in this work, Fiber PROES could be applied to identify and monitor the electron power absorption mode of the plasma during a process. Such plasma operation modes, their stability, and, thus, process-relevant plasma parameters can be affected by a variety of external parameters, such as the position of the matchbox capacitors when the discharge is ignited, the surface conditions of the walls, discharge history, etc. As this ratio is calculated by observing a single emission line, it is not affected by coatings of optical components, which can change the optical transmission of windows. 

Additionally, these experiments show that the electron density and the ion flux do not behave proportionally to each other, which emphasizes the importance of the insight gained by investigating the electron power absorption mode of the plasma.

\ack{This work was supported by the German Research Foundation
in the frame of project T09 of the Collaborative Research Center TR-87 (Grant No. 138690629) and project No. 428942393.}

\bibliographystyle{iopart-num}
\bibliography{bibliography}

\end{document}